# Temperature dependent characteristics of GaSb p-channel MOSFETs with Si-implanted source and drain

Lianfeng Zhao, Zhen Tan, Jing Wang, and Jun Xu[*]

Tsinghua National Laboratory for Information Science and Technology, Institute of Microelectronics, Tsinghua University, Beijing 100084, People's Republic of China

**Abstract**

GaSb p-channel MOSFETs with an atomic layer deposited $Al_2O_3$ gate dielectric and a self-aligned Si implanted source/drain are demonstrated. Thermal anneal conditions are optimized for the source/drain impurity activation. Temperature dependent electrical characteristics are investigated. Different electrical behaviors are observed in two different temperature regions and the mechanisms underneath are proposed. Off-state drain current is generation current dominated in the low temperature regions and is diffusion current dominated in the high temperature regions.

[*]Corresponding author.

Tel.: +86-10-62794752; Fax:+86-10-62771130; E-mail address: junxu@tsinghua.edu.cn (J. Xu).

# 1. Introduction

Unlike most III-V materials with high electron mobility and low hole mobility, such as GaAs and InGaAs [1, 2], GaSb has very high hole mobility, making it attractive as an alternative channel material to silicon, especially for the p-channel MOSFETs [3, 4]. Many efforts have been made for GaSb MOSFET application, such as the source/drain technology [5, 6] and the interface passivation techniques [7-9]. However, technologies for GaSb device application have not been fully developed yet. For example, ion implantation in GaSb has traditionally been a challenge, due to the formation of hillocks and voids [10]. Besides, GaSb has a very low melting point of 712°C, as compared to 1238°C for GaAs and 1414°C for silicon. Consequently, thermal budget must be carefully controlled. However, dopant activation requires relatively high anneal temperature, especially when the substrate doping concentration is high [11]. Furthermore, temperature dependent electrical characteristics of MOSFETs are important, since modern integration circuits usually operate at elevated temperatures due to the power dissipation of the circuit [12]. Although there are some studies on temperature effects on devices based on other materials [13], temperature dependent characteristics of GaSb MOSFETs have not been investigated yet.

In this paper, GaSb p-channel MOSFETs with an atomic layer deposited $Al_2O_3$ gate dielectric and a self-aligned Si ion implanted source/drain are demonstrated. Thermal anneal conditions are carefully optimized. It is found that thermal anneal at 680°C is favorable to improve the performance of GaSb MOSFETs. Temperature dependent electrical characteristics of the fabricated GaSb p-channel MOSFETs are carefully investigated. The fabricated GaSb MOSFETs show different electrical characteristic behaviors in two temperature regions. The mechanism of the off-state drain current are analyzed through the temperature dependent measurements.

## 2. Experiment

GaSb p-channel MOSFETs were fabricated using a self-aligned gate-first flow. Two-inch Te-doped n-type GaSb (100)-oriented wafers with a doping concentration of $4-8\times10^{16}$ cm$^{-3}$ were used as starting substrates for MOSFET fabrication. After initial surface cleaning and passivation by sequential immersion in acetone, ethanol, isopropanol, HCl and (NH$_4$)$_2$S solutions, a 10 nm Al$_2$O$_3$ dielectric layer was deposited by atomic layer deposition (ALD) at 200 °C using a Beneq TFS 200 ALD system. Trimethylaluminum (TMA) and water were used as precursors. Ni/Au gate metal was then deposited and patterned using a lift-off process. The source and drain regions were selectively implanted using Si ion with an implant does of $2\times10^{14}$ cm$^{-2}$ at 30keV. Various thermal anneal conditions (30min forming gas anneal (FGA) at 400°C, 30s rapid thermal anneal (RTA) at 600°C, 650°C, and 680°C) were investigate to optimize the p$^+$/n diode characteristics for the p-channel MOSFETs. After removing the Al$_2$O$_3$ encapsulation layer by KOH solution, the source and drain metals were deposited and patterned by electron beam evaporation of Ni/Au using a lift-off process. Fig. 1 shows the cross-sectional schematic of the p-channel MOSFET structure. Temperature dependent electrical characteristics were recorded using an Agilent B1500A semiconductor device analyzer and a Cascade Summit 11000 AP probe system.

## 3. Results and discussion

A relatively high Si implantation dose was used in this work, due to the relatively high substrate doping concentration. Consequently, a high activation temperature is required to activate the dopants. Various anneal conditions were investigate to optimized the p$^+$/n diode characteristics for the p-channel MOSFETs. Fig. 2 compares the normalized IV characteristics of the p$^+$/n diode annealed at different conditions. Good diode characteristics cannot be obtained with thermal anneal below 650°C, while increasing the thermal anneal temperature to 680°C will lead to a current forward-reverse ratio larger than $10^4$, indicating that thermal anneal at 680°C is favorable for the Si dopant activation.

Fig. 3 (a) shows the transfer characteristics of the fabricated 10 *μm* gate length GaSb p-channel MOSFETs annealed at 680°C. $I_{on}/I_{off}$ ratio as large as ~450 is achieved with drain bias $V_{ds}$ at -1 V. The threshold voltage $V_t$ was determined as -0.325 V from the transfer characteristics measured at $V_{ds}$=-50mV using the linear extrapolation method [14]. The subthreshold swing (SS) of the device at $V_{ds}$=-50mV is ~0.88 V/dec. Fig. 3(b) shows the dc output characteristics of the fabricate GaSb p-channel MOSFETs annealed at 680°C. A maximum drain current $I_{ds}$ of ~1.1mA/mm is obtained at a gate bias $V_{gs}$=-4V and a drain bias $V_{ds}$=-2.5V.

Fig. 4 (a) shows typical transfer characteristics of the fabricated GaSb MOSFETs for temperatures ranging from 240k to 390k at $V_{ds}$ = -1V. The $I_{on}/I_{off}$ ratio decreases as the measurement temperature increases, which is mostly due to the increase of the off-state drain current $I_{off}$ at higher temperatures. In long-channel devices, the off-state drain current is dominated by leakage from the drain-well and well-substrate reverse-bias pn junctions [12]. Fig. 4(b) shows the off-state drain current versus measurement temperature curves. It can be clearly seen that the off-state drain current increases with rising temperature and the lg($I_{off}$) is linear to 1/T with different slopes in two temperature regions. The slope of the curve is greater in temperature regions from 330k to 450k than that in temperature regions from 220k to 330k. Hereafter we refer to these two temperature regions as high temperature regions and low temperature regions, respectively. The different curve slopes in the two temperature regions indicates different leakage current mechanisms. In the high temperature regions, the off-state drain current is diffusion current dominated, and in the low temperature regions, the off-state drain current is generation current dominated. This explanation is confirmed by the temperature dependent measurements and the Shockley-Read-Hall (SRH) generation-recombination theory as below.

According to the Shockley-Read-Hall (SRH) generation-recombination theory [15, 16], the current of the reverse-biased pn junction consists of two components: the diffusion current $J_{dA}$ and the generation current $J_{gA}$, which can be written as follows:

$$J_{dA} = q n_i^2 \frac{D_n}{L_n N_A} \tag{1}$$

$$J_{gA} = \frac{q n_i W_A}{\tau_g} \tag{2}$$

where q is elementary charge, $n_i$ the intrinsic carrier concentration, $D_n$ the diffusion coefficient of electrons in the p side, $W_A$ the volume depletion width, $\tau_r$ the recombination lifetime, $\tau_g$ the generation lifetime, and $L_n = \sqrt{D_n \tau_r}$ is the electron diffusion length.

Given $n_i \propto \exp\left(-\frac{E_g}{2kT}\right)$ and $\tau_g = \tau_r \exp\left(\frac{|E_T - E_i|}{kT}\right)$, $J_{dA}$ and $J_{gA}$ should be given by

$$J_{dA} \propto \exp\left(-\frac{E_g}{kT}\right) \tag{3}$$

$$J_{gA} \propto \exp\left(-\frac{E_T}{kT}\right) \tag{4}$$

where $E_g$ is the band gap and $E_t$ is the trap energy level. Since $E_t$ is smaller than $E_g$, the slope of an Arrhenius plot of $I_{off}$ versus $1/kT$ should be larger when $I_{off}$ is diffusion current dominated and should be smaller when $I_{off}$ is generation current dominated.

Fig 4(c) and (d) plots the $\ln(I_{off})$ versus $1/kT$ curves in the high temperature regions and low temperature regions, respectively. The slope of the curves are extracted as -0.52 and -0.252 respectively, within reasonable standard errors. The larger slope of the curve in the high temperature regions than that of the curve in the low temperature regions confirms that the off-state drain current is diffusion current dominated in high temperature regions and is generation current dominated in low temperature regions. This phenomenon observed in GaSb is consistent with previous studies in Si pn junctions [17]. The slope of -0.52 in the high temperature regions is smaller than the band-gap $E_g$ of GaSb, which is due to the simultaneous influence of the generation current component.

## 4. Conclusion

We have fabricated and studied GaSb p-channel MOSFETs with an atomic layer deposited $Al_2O_3$ gate dielectric and a self-aligned Si ion implanted source/drain. Various thermal anneal conditions were investigated and found that samples treated with 30s RTA at 680°C shows the best performance than the other samples. The mechanism of the off-state drain current was analyzed through the temperature dependent characteristics of the fabricated GaSb p-channel MOSFETs. It is found that the off-state drain current is generation current dominated in low temperature regions and is diffusion current dominated in high temperature regions.

**Acknowledgments**

This work was supported in part by the State Key Development Program for Basic Research of China (No. 2011CBA00602), and the National Science and Technology Major Project (No. 2011ZX02708-002).

**Figure Captions**

Fig. 1. Cross-sectional schematic of the fabricated GaSb p-channel MOSFET structure.

Fig. 2. Diode characteristics of the $p^+$/n junction with various thermal anneal conditions.

Fig. 3. (a) Transfer ($I_{ds}$-$V_{gs}$) characteristics of the fabricated GaSb p-channel MOSFETs. (b) Output ($I_{ds}$-$V_{ds}$) characteristics of the fabricated GaSb p-channel MOSFETs.

Fig. 4. (a) Temperature dependent transfer ($I_{ds}$-$V_{gs}$) characteristics of the fabricated GaSb p-channel MOSFETs. (b) Temperature dependent off-state drain current characteristics. (c) Off-state drain current in the high temperature regions. (d) Off-state drain current in the low temperature regions.

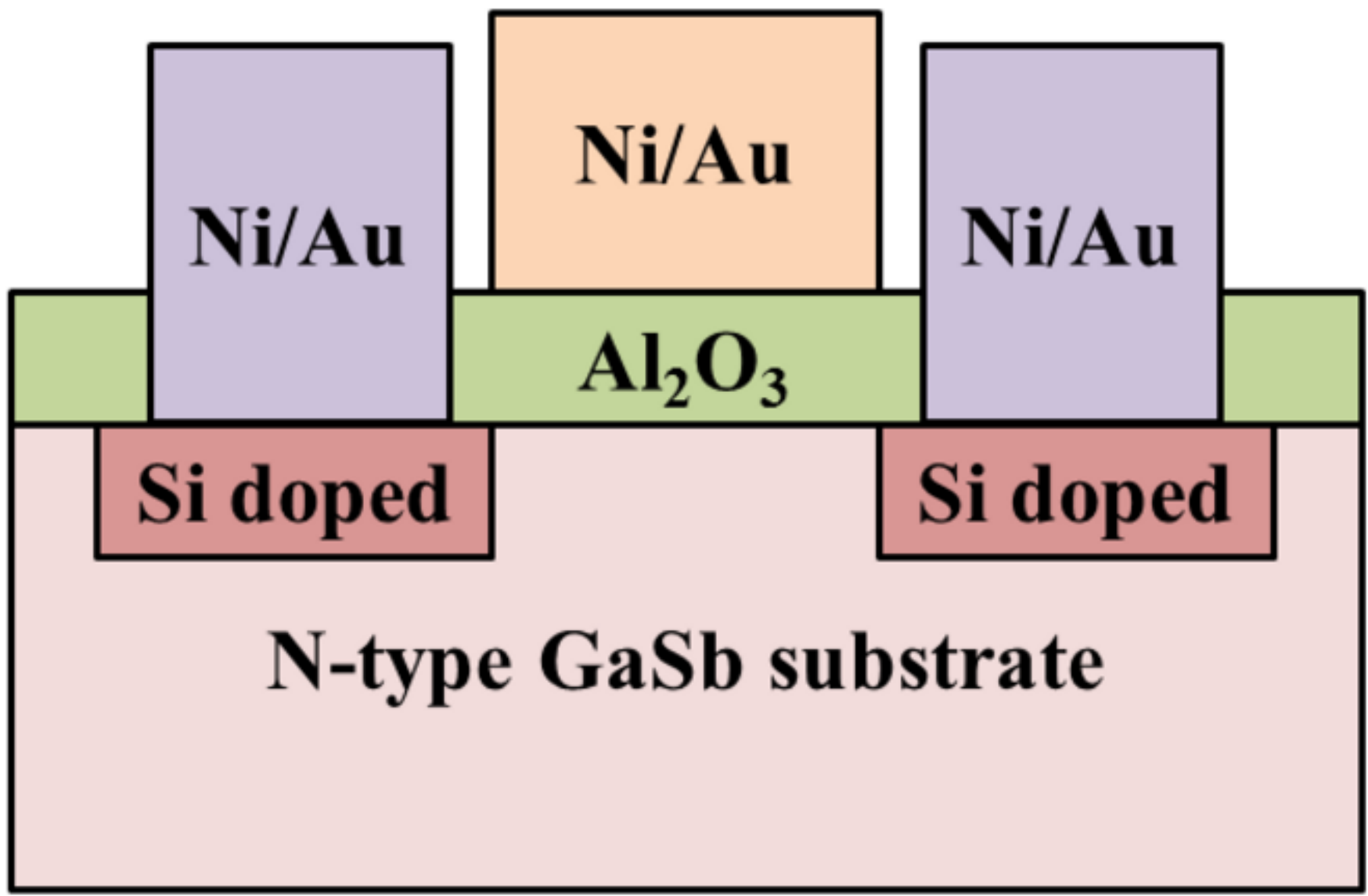

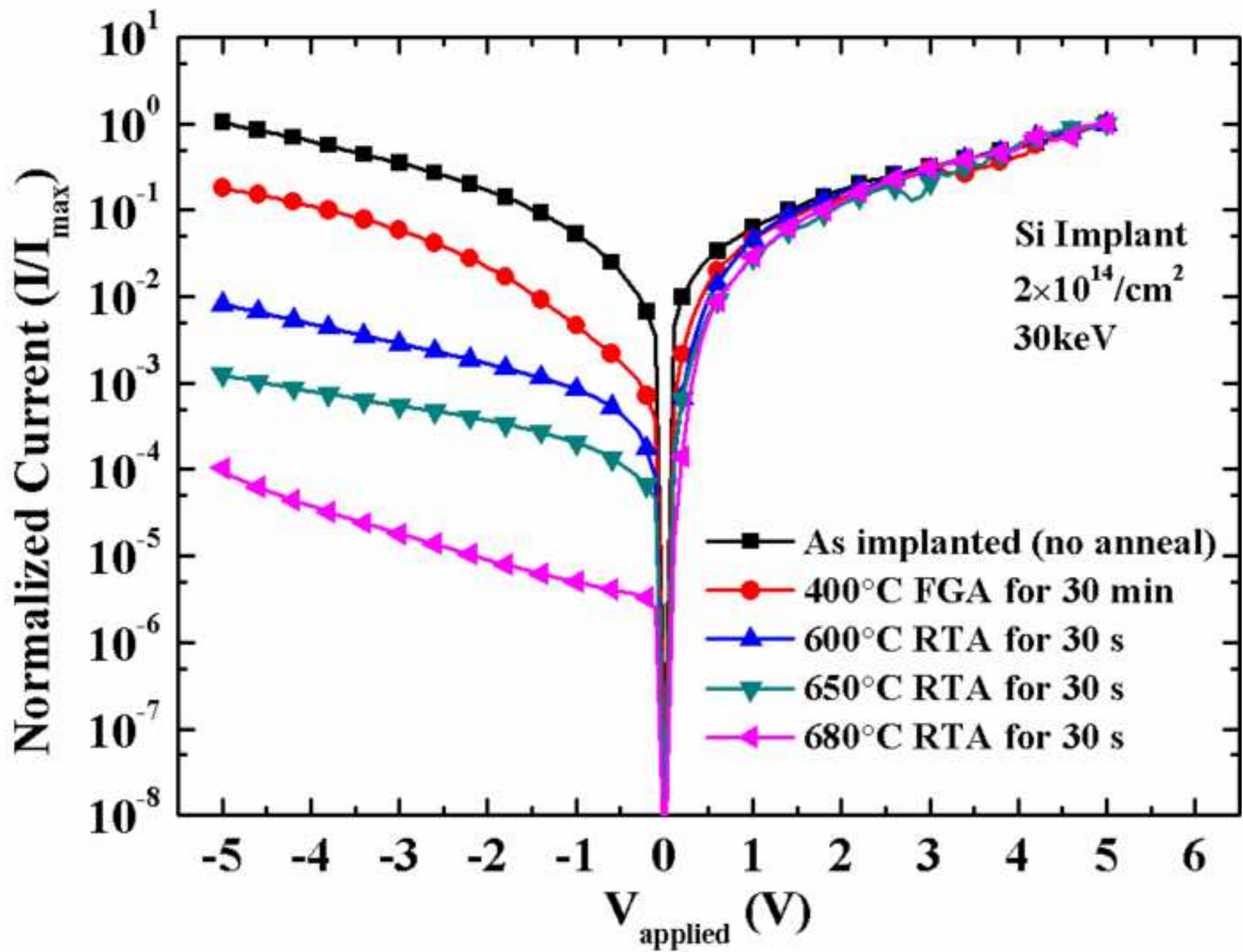

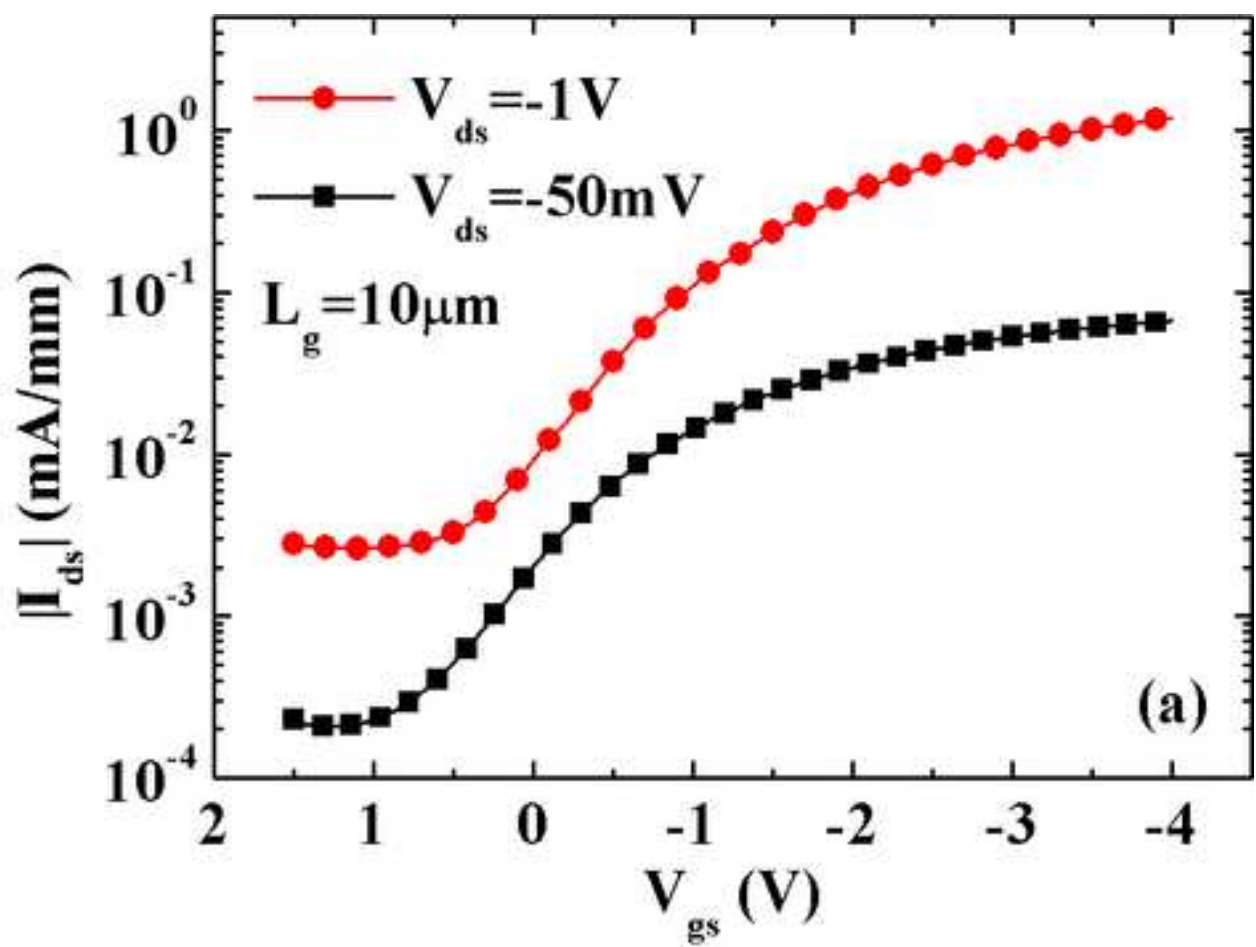
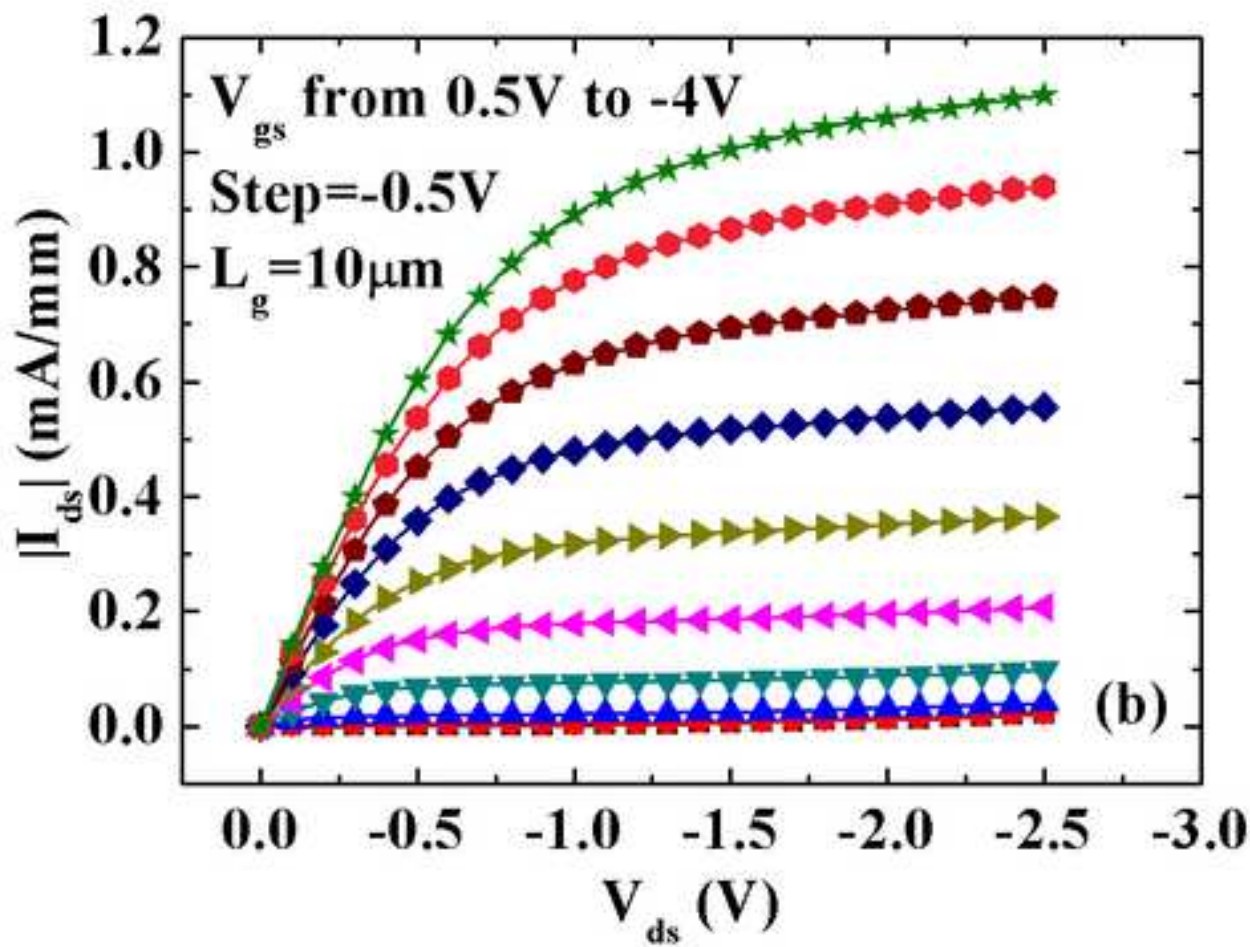

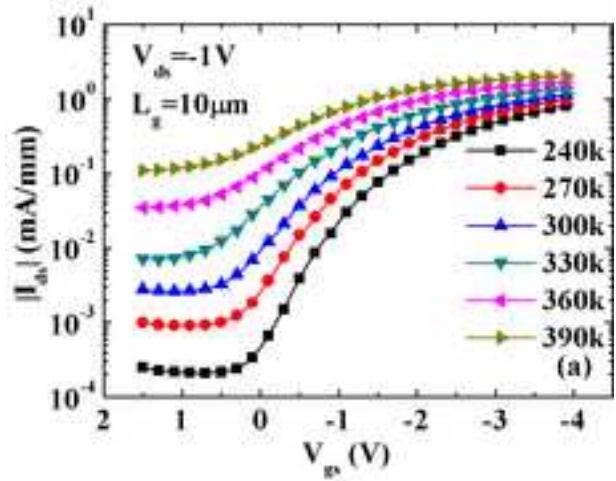
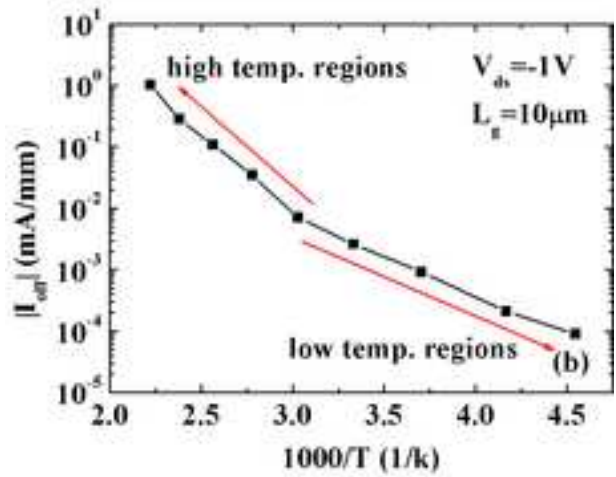
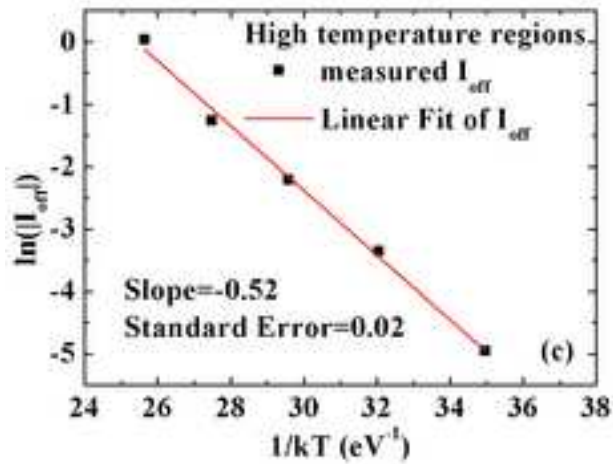
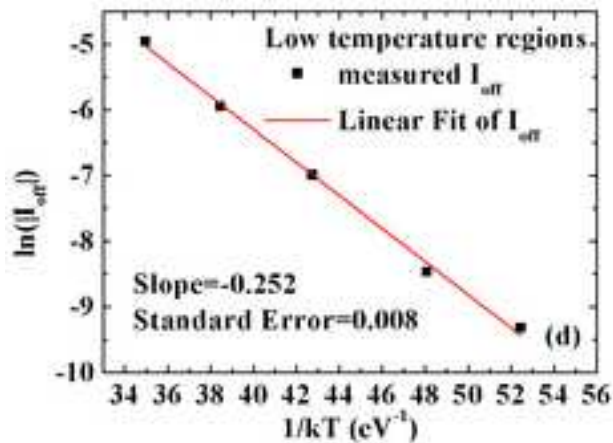